\newcommand{\diracslash}[1]{#1\llap{/\kern2pt}}

\newcommand{\be}{\begin{equation}}
\newcommand{\ee}{\end{equation}}
\newcommand{\bea}{\begin{eqnarray}}
\newcommand{\eea}{\end{eqnarray}}
\newcommand{\ba}[1]{\begin{array}{#1}}
\newcommand{\ea}{\end{array}}

\documentclass[onecolumn,prd,showpacs,preprintnumbers,amsmath,amssymb,aps,float,floatfix]{revtex4-1}
\usepackage{graphicx}
\usepackage{dcolumn}
\usepackage{bm}
\begin{document}
\preprint{}
\title{Cavitation and thermal dilepton production in QGP}
\author{Jitesh R. Bhatt{\footnote {email: jeet@prl.res.in}},
Hiranmaya Mishra{\footnote {email: hm@prl.res.in}} and
V. Sreekanth{\footnote {email: skv@prl.res.in}}}
\affiliation {Theoretical Physics Division, Physical Research Laboratory, Navrangpura, Ahmedabad - 380 009, India}
\date{\today}
\def\be{\begin{equation}}
\def\ee{\end{equation}}
\def\bearr{\begin{eqnarray}}
\def\eearr{\end{eqnarray}}
\def\zbf#1{{\bf {#1}}}
\def\bfm#1{\mbox{\boldmath $#1$}}
\def\hf{\frac{1}{2}}
\begin{abstract}
We study the \textit{non-ideal} effects arising due to viscosity (both bulk and shear), 
equation of state ($\varepsilon\neq 3P$) 
and cavitation on thermal dilepton production from QGP at RHIC energies. 
We calculate the first order corrections to the dilepton production rates due to 
shear and bulk viscosities. Ignoring the cavitation can lead to a wrong estimation 
of dilepton spectra. 
We show that the shear viscosity can enhance the thermal dilepton spectra whereas the 
bulk viscosity can suppress it. 
We present the combined effect of bulk and shear viscosities on the dilepton spectra. 

\end{abstract}
 \maketitle

\section{INTRODUCTION}
\label{intro}
\par
QGP formed in the relativistic heavy ion collider (RHIC) experiments is considered as the most perfect fluid in the nature
\cite{Hirano:2005wx,Ollitrault:2006va,Schf-Teaney-09R}. 
Experiments point towards a very low value for the shear-viscosity to entropy density $\eta/s\sim1/4\pi$- 
the KSS limit\cite{Kovtun:2004de}, 
for this strongly coupled matter formed at RHIC\cite{RHIC}. Generally the effect of bulk-viscosity $\zeta$ is neglected
for a system obeying a relativistic equation of state (EoS). This is because 
$\zeta$ scales like $c_s^2-\frac{1}{3}$, where $c_s^2$ is the speed of sound. Thus 
for a system with \textit{ideal} EoS ($\varepsilon \approx 3P$), where $\varepsilon$ is the energy density and 
$P$ is the pressure ($\frac{P}{\varepsilon}=1/3=c_s^2$), 
the effect $\zeta$ is neglected\cite{wein}. However in the temperature regime close to the the critical 
temperature $T_c$, the bulk viscosity may not be a negligible quantity. In fact, the  
recent lattice QCD results indicate that the quark-gluon matter EoS departures 
from the \textit{ideal} EoS description near critical temperature $T_c$\cite{Bazavov:2009zn}. 
The ratio of the bulk viscosity to the entropy density $s$
calculated from the lattice data indicate $\zeta/s$ too 
has a peak and $\zeta>>\eta$ at the regime $T\approx T_c$ 
\cite{Meyer:2007dy,Karsch:2007jc,Kharzeev:2007wb,Moore:2008ws,CaronHuot:2009ns,Romatschke:2009ng}.
Second order relativistic hydrodynamics models have been used to describe the role of viscosities dynamics on
the evolution of the system
\cite{Baier:2006gy,Romatschke:2007mq,Song:2007ux,Luzum:2008cw,Molnar:2008xj,Song:2008hj,Luzum:2009sb}. 
Using one dimensional hydrodynamics role of bulk viscosity in heavy ion collisions is recently analysed\cite{Fries:2008ts,
Rajagopal:2009yw}. 
Incorporation of bulk viscosity into heavy ion scenario can bring in interesting phenomena of \textit{cavitation}. 
Since bulk (shear) viscosity reduces the longitudinal pressure of the system, during the 
evolution of the system with sufficient values of the viscosities, effective pressure of the fluid can become zero causing cavitation. 
Cavitation leads the fluid to break apart into fragments, making further hydrodynamical description invalid
\cite{Rajagopal:2009yw,Torrieri:2007fb,Torrieri:2008ip}.
In addition the viscous effects can modify 
the temperature profile and thereby it can change the particle distribution
functions of the plasma\cite{deGroot}. 
Using kinetic theory methods one can include these corrections in the distribution functions and this may have observable 
consequences in the observables\cite{Teaney:2003kp,Dusling:2007gi,Monnai:2009ad,Denicol:2009am}. 

Thermal photons and dileptons are among the most promising probes of the hot and dense matter created in relativistic heavy ion 
collisions\cite{Feinberg:1976ua,Shuryak:1978ij}. 
As their mean free path is larger than the transverse size of the fireball, they can escape from the system and there by provide 
information about the thermodynamic state and space-time history of the matter created in heavy ion collisions\cite{kapusta}. 
Production rates of these probes (particles) depend on the temperature of the system and by knowing the appropriate initial conditions 
time evolution of the temperature of the system can be obtained by using the equations of the hydrodynamics. 
Once  the temperature profile is obtained, the calculation of the thermal spectra can be done 
by  evaluating the  cross-section of the underlying scattering processes.
We refer readers Refs.\cite{Alam:1996fd,Alam:1999sc,Gale:2003iz,Peitzmann:2001mz,vanHees:2007th} 
for excellent reviews on the subject. 
Thermal photon production have been studied under various conditions by several authors
\cite{Kapusta:1991qp,Ruuskanen:1991au,Baier:1991em,Thoma:1994fd,Srivastava:2001dn,Arnold:2001ba}. 
Thermal photons from quark-gluon plasma (QGP) in the presence of shear viscosity was studied recently in 
Refs.\cite{Bhatt:2009zg,Dusling:2009bc,PeraltaRamos:2010cw} and they were proposed as a tool to measure 
the shear viscosity of the matter formed in the heavy ion collisions\cite{Bhatt:2009zg,Dusling:2009bc}. 
As argued earlier in the temperature regime $T\geqq T_c $, the finite bulk viscosity may
significantly influence the hydrodynamic evolution of the system\cite{Fries:2008ts,Rajagopal:2009yw}.
Recently the role of these \textit{non-ideal} effects due to EoS, bulk viscosity and cavitation were considered by us in thermal 
photon production\cite{Bhatt:2010cy}. 
We showed that all these effects can alter the particle spectra in a significant manner. It was seen that if the effect 
of cavitation were not included properly one will end up with erroneous estimates of the particle production rates. 
Thermal dilepton production using the equations of  \textit{ideal} hydrodynamics is well studied by many authors
\cite{McLerran:1984ay,Kajantie:1986cu,Kajantie:1986dh}. 
However, only recently the thermal dilepton-production from QGP in the presence of shear viscosity was studied\cite{Dusling:2008xj}. 
In the present work we investigate the effects of finite bulk-viscosity 
on the  thermal dilepton production from QGP.
The main source of thermal dileptons is from the 
quark-anti-quark annihilations: $q\bar q\rightarrow \gamma^* \rightarrow l^+ l^-$. 
The cross-section of this lowest order $\alpha^2$ process is well known\cite{rvogtBook}. 
There are other higher order processes which may also contribute in thermal dilepton production\cite{Altherr:1992th,Thoma:1997dk}, 
however we are not considering them in this present analysis.  
Thermal dileptons from the annihilation process is dominant in the window of intermediate invariant mass and transverse momentum 
of the lepton pair $1< M,~p_T <3$ GeV\cite{vanHees:2007ma,Akiba:2009es}.

\par

In Ref.\cite{Dusling:2008xj} authors has studied the role of shear viscosity in thermal dilepton spectra. 
However we note that they had used an \textit{ideal} EoS 
for the calculation and the effect of bulk viscosity was not considered. 
It has been shown in our previous work that bulk viscosity plays a dual role in heavy ion collisions, on the one hand it enhances the 
time by which system cools down to $T_c$ and on the other hand it can make the hydrodynamic treatment invalid much before it 
reaches $T_c$. 
\textit{Non-ideal} EoS also makes the system spend more time near $T_c$ thus increasing the thermal particle production
\cite{Bhatt:2010cy}. 
In this work we are studying the effect of \textit{non-ideal} EoS, bulk viscosity and cavitation on the thermal dilepton production 
from the QGP. 
Firstly, by taking the viscous modified distribution functions using the 14-moment Grad's method results, we 
calculate the first order correction due to both bulk and shear viscosity in the dilepton production rate. 
We use a recent lattice QCD calculation result for \textit{non-ideal} EoS. 
We take $\eta/s=1/4\pi$ in our analysis and we use lattice QCD result for $\zeta/s$ following Ref.\cite{Rajagopal:2009yw}. 
By using second order relativistic causal dissipative hydrodynamics we analyse the evolution dynamics of the system. 
Where we treat the expanding system as one dimensional boost invariant flow. Since boost-invariant hydrodynamics leads 
to an underestimation of the effects of bulk viscosity, our dilepton spectra acts as a conservative estimate of the 
effects. 
At early stages of the expansion, transverse flow can be neglected. 
Eventhough we are not including transverse flow, we believe its effects could remain small as cavitation can reduce the 
hydrodynamical evolution.

\section{Thermal Dilepton production rates in QGP}
\label{thermal-dileptons}

In QGP the dominant mechanism for the production of thermal dileptons comes from $q\bar q$ annihilation process 
$q\bar q\rightarrow \gamma^* \rightarrow l^+ l^-$. From kinetic theory rate of dilepton production 
(number of dileptons produced per unit volume per unit time) for this process is given by 
\begin{equation}
 \frac{dN}{d^4x}=\int \frac{d^3\textbf{p}_1}{(2\pi)^3} \frac{d^3\textbf{p}_2}{(2\pi)^3} f(E_1,T) f(E_2,T)
 v_{rel} ~g^2\sigma(M^2),
\end{equation}
where $p_{1,2}=(E_{1,2},\textbf{p}_{1,2})$ is the four momentum of quark or anti-quark with  
$E_{1,2}=\sqrt{\textbf{p}_{1,2}^2+m_q^2}\simeq\rvert\textbf{p}_{1,2}\lvert$ neglecting the quark masses. 
Here $M^2=(E_1+E_2)^2-(\textbf{p}_1+\textbf{p}_2)^2$ is the invariant mass of the virtual photon. 
The function $f(E,T)=1/(1+e^{E/T})$ is the quark (anti-quark) distribution function in thermal equilibrium 
and $g$ is the degeneracy factor.  
$v_{rel}=\sqrt{\frac{M^2(M^2-4m_q^2)}{4E_1^2E_2^2}}\sim \frac{M^2}{2E_1E_2}$ is the 
relative velocity of the quark-anti-quark pair and $\sigma(M^2)$ is the thermal dilepton production cross section. 
The cross-section $\sigma(M^2)$ in the Born approximation is 
well known: $g^2\sigma(M^2)=\frac{16\pi\alpha^2\left(\sum_f e_f^2\right)N_c}{3M^2}$ and with $N_f$=2 and $N_c=3$, 
we have $M^2g^2\sigma(M^2)=\frac{80\pi}{9}\alpha^2$\cite{Alam:1996fd}.
Since we are interested in the rate for a given dilepton mass and momentum, we write

\begin{equation}
 \frac{dN}{d^4xd^4p}=\int \frac{d^3\textbf{p}_1}{(2\pi)^3} \frac{d^3\textbf{p}_2}{(2\pi)^3} f(E_1,T) f(E_2,T)
  ~\frac{M^2g^2\sigma(M^2)}{2E_1E_2} \delta^4(p-p_1-p_2) \label{dilrateEq}
\end{equation}
where $p=(p_0=E_1+E_2,\textbf{p}=\textbf{p}_1+\textbf{p}_2)$ is the four momentum of the dileptons. At the present case 
we are interested in the invariant masses larger compared to the temperature i.e.; $M\gg T$, in this limit we can replace 
Fermi-Dirac distribution with classical Maxwell-Boltzmann distribution,
\begin{equation}\label{MxBzn}
 f(E,T) \rightarrow f_0 = e^{-E/T}.
\end{equation}

\section{Viscous corrections to the dilepton production rates}
\label{visc-corr}

Viscosity effects modify the particle production spectra in two ways. Firstly it modifies the temperature and secondly through the 
corrections in distribution functions\cite{Bhatt:2010cy}. 
The first effect effect is incorporated when we calculate the temperature as a function of time by solving dissipative hydrodynamics. 
This will be done in the next section. Once we include bulk viscosity, a novel phenomena like \textit{cavitation} can arise which can 
alter particle production spectra considerably\cite{Bhatt:2010cy}. In the present section we concentrate on the second effect. 
To calculate the viscous modifications to the distribution function 
as a function of the momentum, we need to use the techniques of relativistic kinetic theory\cite{deGroot}. 
Let us write the modified distribution function as $f=f_0+\delta f$, with viscous correction $\delta f=\delta f_\eta + \delta f_\zeta$, 
where $\delta f_\eta$ and $\delta f_\zeta$ represent change 
in the distribution function due to shear and bulk viscosity respectively. 
We can calculate these corrections using 14-moment Grad's method as done in Refs.\cite{Teaney:2003kp,Dusling:2007gi,Bhatt:2010cy}. 
Now modified distribution function from corrections 
due to shear ($\eta$) and bulk ($\zeta$) viscosity (up to quadratic order of momentum) are given by\cite{Bhatt:2010cy}

\begin{equation}
 f(p) = f_0(p) \bigg(1 + \frac{\eta/s}{2 T^3} p^{\alpha}p^{\beta} 
    \nabla_{\langle\alpha} u_{\beta\rangle} 
+ \frac{2}{5}\frac{\zeta/s}{2 T^3} p^{\alpha}p^{\beta}\Delta_{\alpha\beta}\Theta\bigg).\label{delF}
\end{equation}
where $s$ is the entropy density and the operators are defined as 
$\Delta^{\alpha\beta}=g^{\alpha\beta}-u^\alpha\,u^\beta$, $\nabla_{\alpha}=\Delta_{\alpha\beta}\partial^{\beta}$,
$\Theta\equiv\nabla_{\alpha} u^{\alpha}$ and $\nabla_{\langle\alpha} u_{\beta\rangle}=\nabla_{\alpha}\,u_{\beta} 
+ \nabla_{\beta}\,u_{\alpha}
-\frac{2}{3}\,\Delta_{\alpha\beta}\Theta$. 

In order to compute the effect of viscosity on the production rate, we substitute equation (\ref{delF}) 
in dilepton rate equation (\ref{dilrateEq}). 
Thus keeping terms up to the second order in $\eta/s~\rm{and}~\zeta/s$, the dilepton production rates can be written as, 

\begin{eqnarray}
 \frac{dN}{d^4xd^4p}
= \frac{dN^{(0)}}{d^4xd^4p} + \frac{dN^{(\eta)}}{d^4xd^4p} + \frac{dN^{(\zeta)}}{d^4xd^4p}, \label{dilrate}
\end{eqnarray}
with
\begin{eqnarray}
 \frac{dN^{(0)}}{d^4xd^4p}&=&\int \frac{d^3\textbf{p}_1}{(2\pi)^3} \frac{d^3\textbf{p}_2}{(2\pi)^3}~ e^{-(E_1+E_2)/T}
\frac{M^2g^2\sigma(M^2)}{2E_1E_2} \delta^4(p-p_1-p_2) \label{dN0}\\ 
\frac{dN^{(\eta)}}{d^4xd^4p}&=&\int \frac{d^3\textbf{p}_1}{(2\pi)^3} \frac{d^3\textbf{p}_2}{(2\pi)^3}~ e^{-(E_1+E_2)/T}
\left[\frac{\eta/s}{T^3} p_1^{\alpha}p_1^{\beta} \nabla_{\langle\alpha} u_{\beta\rangle} \right]
\frac{M^2g^2\sigma(M^2)}{2E_1E_2} \delta^4(p-p_1-p_2)\label{dNsh}\\
\frac{dN^{(\zeta)}}{d^4xd^4p}&=&\int \frac{d^3\textbf{p}_1}{(2\pi)^3} \frac{d^3\textbf{p}_2}{(2\pi)^3}~ e^{-(E_1+E_2)/T}
\left[\frac{2}{5}\frac{\zeta/s}{T^3} p_1^{\alpha}p_1^{\beta}\Delta_{\alpha\beta}\Theta \right]
\frac{M^2g^2\sigma(M^2)}{2E_1E_2} \delta^4(p-p_1-p_2).\label{dNblk}
\end{eqnarray}
The first term (given by equation (\ref{dN0})) is the one without any viscous corrections (\textit{ideal} part) and 
is well known\cite{rvogtBook}
\begin{equation}
  \frac{dN^{(0)}}{d^4xd^4p}=\frac{1}{2}~\frac{M^2g^2\sigma(M^2)}{(2\pi)^5}~e^{-p_0/T}.\label{dilrate0}
\end{equation}
The first order correction to the rate due to shear viscosity- given by equation (\ref{dNsh}), is calculated in Ref.
\cite{Dusling:2008xj} 
and the final expression is  
\begin{equation}
\frac{dN^{(\eta)}}{d^4xd^4p}=\frac{1}{2}~\frac{M^2g^2\sigma(M^2)}{(2\pi)^5}~e^{-p_0/T}
~\frac{2}{3}\left[\frac{\eta/s}{2T^3} p^{\alpha}p^{\beta} \nabla_{\langle\alpha} u_{\beta\rangle} \right].\label{dilrate1sh}
\end{equation}
Let us next proceed to estimate the correction to the rate due to bulk viscosity from equation (\ref{dNblk}). We can write 
\begin{equation}
 \frac{dN^{(\zeta)}}{d^4xd^4p}=\int \frac{d^3\textbf{p}_1}{(2\pi)^6} ~ e^{-(E_1+E_2)/T}
\left[\frac{2}{5}\frac{\zeta/s}{T^3} p_1^{\alpha}p_1^{\beta}\Delta_{\alpha\beta}\Theta \right]
\frac{M^2g^2\sigma(M^2)}{2E_1E_2} \delta(p_0-E_1-E_2)=\frac{2}{5}\frac{\zeta/s}{T^3} I^{\alpha\beta}(p)\Delta_{\alpha\beta}\Theta,
\end{equation}
where we have represented
\begin{equation}
I^{\alpha\beta}=\int \frac{d^3\textbf{p}_1}{(2\pi)^6} ~ e^{-(E_1+E_2)/T}
 p_1^{\alpha}p_1^{\beta} \frac{M^2g^2\sigma(M^2)}{2E_1E_2} \delta(p_0-E_1-E_2)\label{Iform}
\end{equation}
Now we write the second rank tensor $I^{\alpha\beta}$ in the most general form constructed out of $u^{\alpha}$ and $p^{\alpha}$:
\begin{equation}
 I^{\alpha\beta}=a_0 g^{\alpha\beta}+a_1 u^\alpha u^\beta+a_2 p^\alpha p^\beta + a_3(u^\alpha p^\beta + u^\beta p^\alpha)\,.
\end{equation}
Note that because of the identity $u^{\alpha}\Delta_{\alpha\beta}=0$, the coefficients of $I^{\alpha\beta}$ which are going 
to survive after contraction with $\Delta_{\alpha\beta}$ are $a_0$ and $a_2$. We construct two projection operators to get these 
coefficients, i.e.; $Q^1_{\alpha\beta}I^{\alpha\beta}=a_0$ and $Q^2_{\alpha\beta}I^{\alpha\beta}=a_2$, so that 
\begin{equation}
  \frac{dN^{(\zeta)}}{d^4xd^4p}=\frac{2}{5}\frac{\zeta/s}{T^3} \left[\left(Q^1_{\mu\nu}I^{\mu\nu}\right)g^{\alpha\beta}
+\left(Q^2_{\mu\nu}I^{\mu\nu}\right)p^\alpha p^\beta \right] \Delta_{\alpha\beta}\Theta.
\end{equation}
The expressions for the projection operators in the local rest frame of the the medium ($u^\alpha=(1,\bar 0)$) are
\begin{eqnarray}
Q^1_{\alpha\beta}&=&\frac{1}{2|{\bf p}|^2}[|{\bf p}|^2g_{\alpha\beta}+M^2u_{\alpha}u_{\beta}
+p_{\alpha}p_{\beta}-2p_0u_{\alpha}p_{\beta}]\,,\\
Q^2_{\alpha\beta}&=&\frac{1}{2|{\bf p}|^4}[|{\bf p}|^2g_{\alpha\beta}+(3p_0^2-|{\bf p}|^2)u_{\alpha}u_{\beta}
+3p_{\alpha}p_{\beta}-6p_0u_{\alpha}p_{\beta}]\,.
\end{eqnarray}
With the help of definition of $I^{\alpha\beta}$ i.e.; equation (\ref{Iform}), we can calculate 
$\left(Q^1_{\mu\nu}I^{\mu\nu}\right)$ and $\left(Q^2_{\mu\nu}I^{\mu\nu}\right)$. We now write the final expression for the 
first order correction due to bulk viscosity in dilepton rate
\begin{equation}
\frac{dN^{(\zeta)}}{d^4xd^4p}
=\frac{1}{2}~\frac{M^2g^2\sigma(M^2)}{(2\pi)^5}~e^{-p_0/T} 
\left[\frac{2}{3}\left( \frac{2}{5}\frac{\zeta/s}{2T^3}p^\alpha p^\beta \Delta_{\alpha\beta}\Theta \right) 
-\frac{2}{5}\frac{\zeta/s}{4T^3} M^2\Theta \right],\label{dilrate1bv}
\end{equation}
where we have used the identity $\Delta^\alpha_\alpha=3$. 
\par
The \textit{total} dilepton rate, including the first order viscous corrections due to both shear and bulk viscosity is obtained 
by adding equations (\ref{dilrate0}), (\ref{dilrate1sh}) and (\ref{dilrate1bv}). 

Apart from rates as function of four momentum of the dileptons we will be interested in particle production 
as a function of invariant mass ($M$), transverse momentum ($p_T$) and rapidity ($y$) of the dilepton pair. This can be obtained 
from equation (\ref{dilrate}) by changing the variables appropriately
\cite{Alam:1996fd},
\begin{eqnarray}
 \frac{dN}{d^4xdM^2d^2p_Tdy}&=&\frac{1}{2} \frac{dN}{d^4xd^4p}\nonumber\\ 
&=&\frac{1~}{2^3}\frac{5\alpha^2}{9\pi^4}~e^{-p_0/T}\left[1+
\frac{2}{3}\left( \frac{\eta/s}{2T^3} p^{\alpha}p^{\beta} \nabla_{\langle\alpha} u_{\beta\rangle} + 
\frac{2}{5}\frac{\zeta/s}{2T^3}p^\alpha p^\beta \Delta_{\alpha\beta}\Theta \right) 
-\frac{2}{5}\frac{\zeta/s}{4T^3} M^2\Theta \right].\label{dlratefinal}
\end{eqnarray}

\section{Viscous hydrodynamics and cavitation}
\label{hydro}
In order to study the dilepton production from the QGP formed in heavy ion collision we need to understand the evolution dynamics 
of the system. By obtaining the temporal distribution of temperature and information about viscosity coefficients we can study the 
dilepton spectra. 

We study the QGP formed in high energy nuclear collisions using causal dissipative second 
order hydrodynamics of Israel-Stewart\cite{Israel:1979wp}, with the expanding fireball 
treated as having one dimensional boost invariant expanding flow\cite{Bjorken:1982qr}. 
We use the parametrization of the coordinates $t=\tau$ cosh$\,\eta_s$ and $z=\tau$ sinh$\,\eta_s$, 
with the proper time $\tau = \sqrt{t^2-z^2}$ and space-time 
rapidity $\eta_s=\frac{1}{2}\,ln[\frac{t+z}{t-z}]$. Now the four velocity can be written as 
\begin{equation}\label{4vel}
u^\mu=(\rm{cosh}\,\eta_s,0,0,\rm{sinh}\,\eta_s).
\end{equation}

With this second order theory (For more details on this theory and its application to relativistic heavy ion collisions 
we refer \cite{Muronga:2004sf,Romatschke:2009im,Bhatt:2010cy})
the equations dictating the longitudinal expansion of the medium are given by
\cite{Muronga:2001zk,Heinz:2005zi,Baier:2006um,Muronga:2006zw}: 
\begin{eqnarray}
   \frac{\partial\varepsilon}{\partial\tau} &=& - \frac{1}{\tau}(\varepsilon 
  + P +\Pi - \Phi) \, ,
  \label{evo1} \\
  \frac{\partial\Phi}{\partial\tau} &=& - \frac{\Phi}{\tau_{\pi}}+\frac{2}{3}\frac{1}{\beta_2\tau}
  -\frac{1}{\tau_{\pi}}\left[ \frac{4\tau_{\pi}}{3\tau}\Phi +\frac{\lambda_1}{2\eta^2}\Phi^2\right] 
  \, , 
  \label{evo2} \\
   \frac{\partial\Pi}{\partial\tau} &=& - \frac{\Pi}{\tau_{\Pi}} - \frac{1}{\beta_0\tau} .
\label{evo3}
\end{eqnarray}
The effects due to shear and bulk viscosity are represented via $\Phi$ and $\Pi$ respectively and they alter the 
equilibrium pressure. 
The first equation is the equation of motion and 
the other two equations: (\ref{evo2} $\&$ \ref{evo3}) are 
evolution equations for $\Phi$ and $\Pi$ governed by 
their relaxation times $\tau_{\pi}$ and $\tau_{\Pi}$ respectively. 
The coefficients $\beta_0$ and $\beta_2$ are related with the relaxation time $\tau_\Pi=\zeta\,\beta_0\,\rm{and}\, 
\tau_\pi=2\eta\,\beta_2.$
We use the $\mathcal N\,=\,4$ supersymmetric Yang-Mills theory expressions for $\tau_\pi$ and $\lambda_1$
\cite{Natsuume:2007ty,Baier:2007ix,Bhattacharyya:2008jc}: $\tau_{\pi} = \frac{2-\ln 2}{2\pi T}$ and $\lambda_1 = \frac{\eta}{2\pi T}$ and 
we take $\tau_\pi(T)=\tau_\Pi(T)$ following Ref.\cite{Fries:2008ts}. 

Apart from these three equations (\ref{evo1} - \ref{evo3}), we need to provide the EoS to study the hydrodynamical evolution 
of the system. 
We use the recent lattice QCD result of A. Bazavov $\it{et ~al.}$\cite{Bazavov:2009zn} for equilibrium equation of state 
(EoS) (\textit{non-ideal}: $\varepsilon-3P\neq 0$), which becomes significantly important near the critical temperature. 
Parametrised form of their result for trace anomaly is given by
\begin{equation}
\frac{\varepsilon-3P}{T^4} = \left(1-\frac{1}{\left[1+\exp\left(\frac{T-c_1}{c_2}\right)\right]^2}\right)
\left(\frac{d_2}{T^2}+\frac{d_4}{T^4}\right)\ ,
\label{e3pt4}
\end{equation}
where values of the coefficients are $d_2= 0.24$~GeV$^2$, $d_4=0.0054$~GeV$^4$, $c_1=0.2073$~GeV, 
and $c_2=0.0172$~GeV. 
The functional form of the pressure is given by \cite{Bazavov:2009zn}
\begin{equation}
\frac{P(T)}{T^4} - \frac{P(T_0)}{T_0^4} = \int_{T_0}^T dT' \,\frac{\varepsilon-3P}{T'^5}\ ,
\label{pt4}
\end{equation}
with $T_0~$= 50 MeV and $P(T_0)$ = 0 \cite{Rajagopal:2009yw}. 
From equations (\ref{e3pt4}) and (\ref{pt4}) we get $\varepsilon$ and $P$ in terms of $T$.
A crossover from QGP to hadron gas around the temperature 200-180 MeV is predicted by this model. 
Throughout the analysis we keep the critical temperature $T_c$ to be 190 MeV. 

Now we need to specify the viscosity prescriptions used in the hydrodynamical model. 
We use recent lattice QCD calculation results of Meyer\cite{Meyer:2007dy}, for determining $\zeta/s$. 
His result indicate the existence a peak of $\zeta/s$ near $T_c$, although the height 
and width of this curve are not well understood. 
We use the parametrization of Meyer's result given in Ref.\cite{Rajagopal:2009yw}: 
\begin{equation}
\frac{\zeta}{s} = a \exp\left( \frac{T_c - T}{\Delta T} \right) + b \left(\frac{T_c}{T}\right)^2\quad{\rm for}\ T>T_c,
\label{zetabys}
\end{equation}
where the parameter $a=\rm{0.901}$ controls the height, $\Delta T=T_c/14.5$ controls the width of the $\zeta/s$ curve 
and $b$ = 0.061.
We take the conservative lower bound of the shear viscosity to entropy density ratio 
$\eta/s=1/4\pi$
in our calculations\cite{Kovtun:2004de}. It is observed that the \textit{non-ideal} 
EoS deviates from the \textit{ideal} case ($\varepsilon=3P$) significantly around the critical temperature and 
around same temperature $\zeta/s$ starts to dominate over $\eta/s$ significantly\cite{Bhatt:2010cy}. 

Let us observe the longitudinal pressure of the system given by the equation, 
\begin{equation}
 P_{z} = P + \Pi - \Phi.
\end{equation} 
Since the bulk viscosity 
contribution $\Pi<0$ always, $\Pi$ and $\Phi$ together can make $P_z$ negative\cite{Rajagopal:2009yw}. 
(We note that with $\eta/s\sim 1/4\pi $- as suggested by RHIC 
experiments, alone is not sufficient for this condition). The condition $P_z=0$ defines the onset of cavitation and the time 
at which it occurs is called cavitation time denoted as $\tau_c$. 
After the onset of cavitation the fluid breaks apart and the theory of hydrodynamics is no longer valid to describe the system
\cite{Torrieri:2007fb,Torrieri:2008ip,Rajagopal:2009yw}. 
So we can only evolve the hydrodynamics code till $\tau_c$ in case of occurrence of cavitation instead of 
$\tau_f$, where $T(\tau_f)=T_c$\cite{Rajagopal:2009yw}.

\section{Dilepton spectra in Heavy Ion Collision}
\label{dilepton yields}

\begin{figure}
\includegraphics[width=8.6cm]{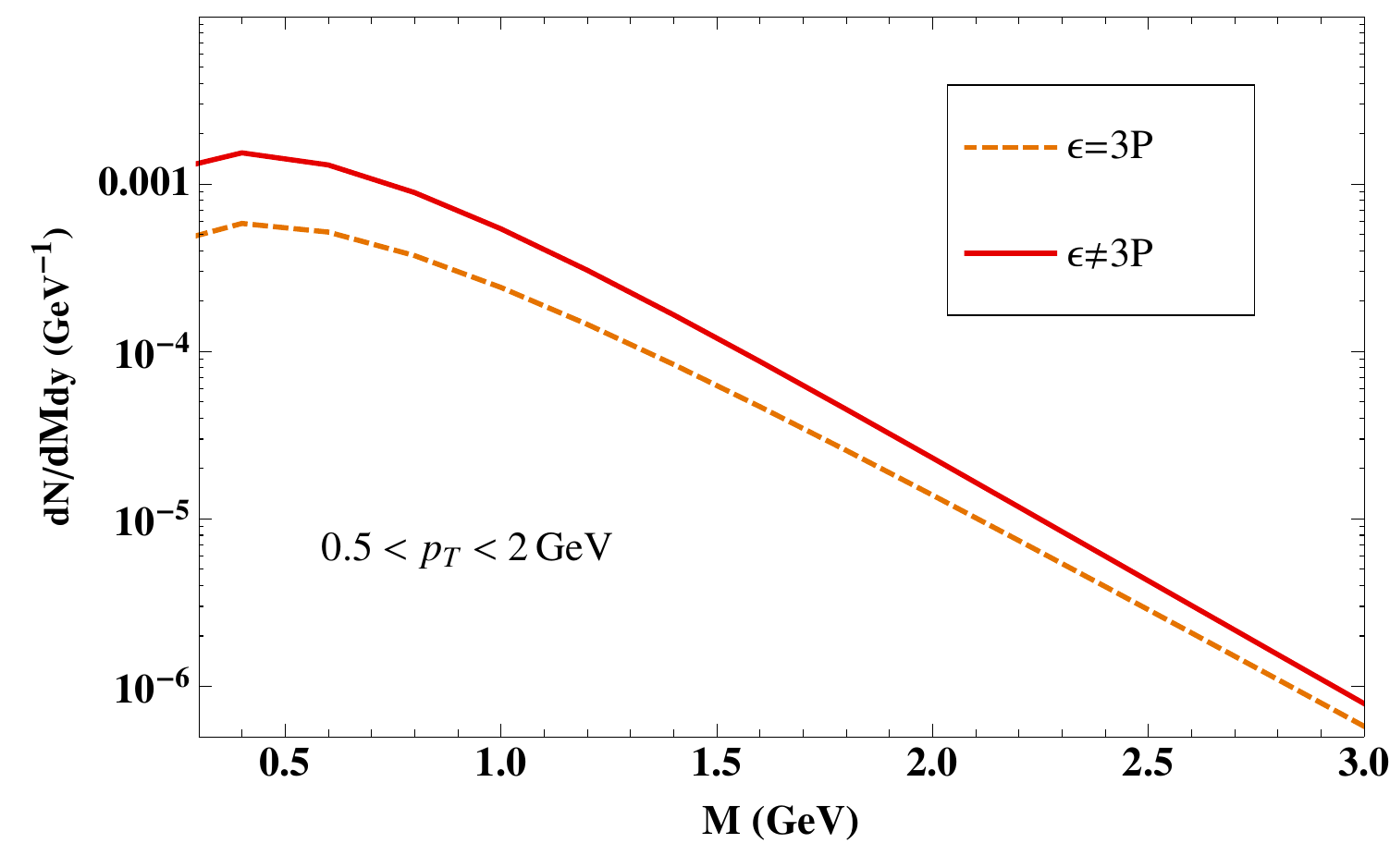}
\caption{Dilepton rate as a function of invariant mass $M$ for different equation of states. Effect of viscosity is not included 
in the hydrodynamical evolution and the distribution functions}\label{fig:2}
\end{figure}

The total dilepton spectrum is obtained by convoluting the dilepton rate with the space-time evolution of the heavy ion collision. 
Dilepton rates are temperature dependent and temperature profile is obtained after hydrodynamically evolving the system as described 
in the previous session. 
In the Bjorken model, the four dimensional volume element is given by 
\begin{equation}
 d^4x=d^2x_Td\eta_s\tau d\tau=\pi R_A^2d\eta_s\tau d\tau,
\end{equation}
where $R_A=1.2 A^{1/3}$ is the radius of the nucleus used for the collision (for $Au,~ A=197$). 
We can calculate different differential rates as functions of $M,~p_T~\rm{and}~y$. In this work we will be calculating the rates 
$dN/(p_Tdp_TdMdy)$ and $dN/dMdy$; and these dilepton yields are obtained from,
\begin{eqnarray}\label{tot-yield}
\left(\frac{dN}{dM^2d^2p_Tdy}\right)_{M,p_T,y}
=\pi R_A^2\int_{\tau_0}^{\tau_1}
d\tau ~\tau \int_{-y_{nuc}}^{y_{nuc}}d\eta_s
\left(\frac{1}{2}\frac{dN}{d^4xd^4p}\right)\nonumber 
\end{eqnarray}
and are given by
\begin{eqnarray}\label{tot-yield1}
\left(\frac{dN}{p_Tdp_TdMdy}\right)_{M,p_T,y} &=&(4\pi M) \pi R_A^2 \int_{\tau_0}^{\tau_1}
d\tau ~\tau \int_{-y_{nuc}}^{y_{nuc}}d\eta_s
\left(\frac{1}{2}\frac{dN}{d^4xd^4p}\right),\\\label{tot-yield2}
\left(\frac{dN}{dMdy}\right)_{M,y} &=&(4\pi M) \pi R_A^2 \int_{\tau_0}^{\tau_1}
d\tau ~\tau \int_{-y_{nuc}}^{y_{nuc}}d\eta_s \int_{p_{T_{min}}}^{p_{T_{max}}}
p_Tdp_T \left(\frac{1}{2}\frac{dN}{d^4xd^4p}\right).
\end{eqnarray}
Here $\tau_0$ and $\tau_1$ are the initial and final values of time that we are interested. 
Generally $\tau_1$ 
is taken as the time taken by the system to reach $T_c$, i.e.; $\tau_f$, 
but in the case of occurrence of cavitation we must set $\tau_1=\tau_c$, the cavitation time, in order to avoid erroneous 
estimation of rates\cite{Bhatt:2010cy}. 
$y_{nuc}$ is the rapidity of the nuclei used for the experiment.

Here we note that the dilepton production rates calculated in Section \ref{visc-corr} correspond to the rest frame of the system. 
So in a longitudinally expanding system, we must replace $f_0$ of equation (\ref{MxBzn}) with $f_0=e^{-u.p/T}$ in equations 
(\ref{tot-yield1}-\ref{tot-yield2}). With the four momentum of the dilepton parametrised as 
$p^{\alpha}$ = $(m_T coshy,p_T cos\phi_p,p_T sin\phi_p,m_Tsinhy)$, where $m_T^2$ = $p_T^2+M^2$\cite{Teaney:2003kp} 
and the four velocity of the medium given by equation (\ref{4vel}) we get, $u.p=m_T cosh(y-\eta_s)$. 
Thus using the 1D boost invariant flow, the factors appearing in the modified rate equation (\ref{dlratefinal}) can be calculated as
\begin{eqnarray}\label{visc-fcts}
 p^{\alpha}p^{\beta} \nabla_{\langle\alpha} u_{\beta\rangle} &=& \frac{2}{3\tau}p_T^2-\frac{4}{3\tau}m_T^2sinh^2(y-\eta_s),\\
 p^\alpha p^\beta \Delta_{\alpha\beta}\Theta &=& -\frac{p_T^2}{\tau}-\frac{m_T^2}{\tau}sinh^2(y-\eta_s), 
\end{eqnarray}
with $\Theta=1/\tau$.

\section{RESULTS AND DISCUSSIONS}

\begin{figure}
\includegraphics[width=8.6cm]{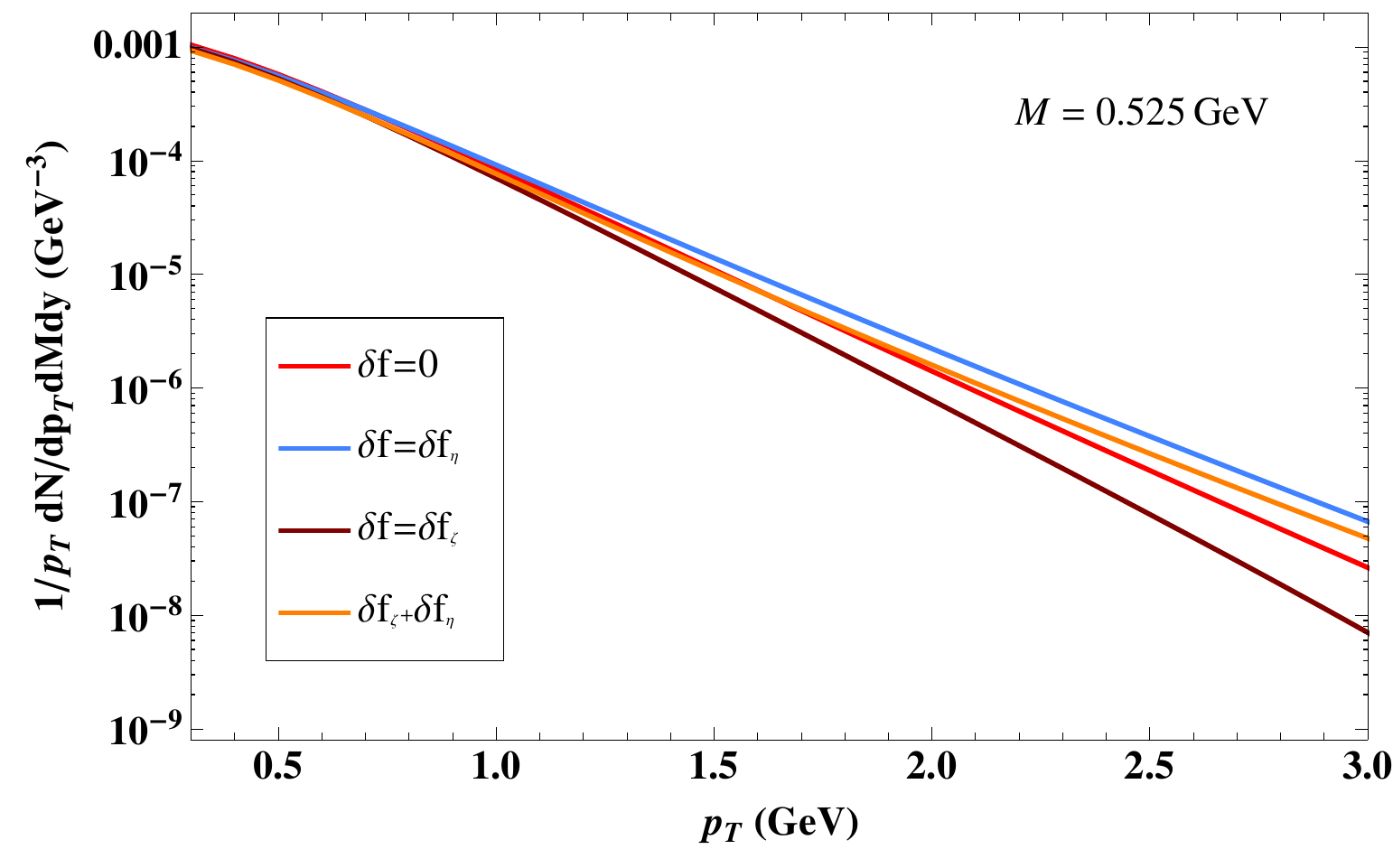}
\caption{Transverse momentum spectra of dileptons from a viscous QGP calculated at $M=0.525~GeV$. 
The solid line shows the dilepton production rate without considering the viscous corrections to the distribution functions. 
The effect of inclusion of viscous corrections due to shear and bulk is shown in separate curves. 
}\label{fig:3}
\end{figure}

We calculate the dilepton yields by obtaining the temperature and bulk viscosity as functions of time by 
solving the hydrodynamical equations with relevant initial conditions. 
For the hydrodynamical evolution of the system we use initial conditions relevant for the RHIC experiment, 
taken from Ref.\cite{Srivastava:2001dn}. Initial time ($\tau_0$) and temperature ($T_0$) are given as 0.5 $fm/c$ and 
310 $MeV$ respectively, whereas $y_{nuc} = 5.3$. 
The initial values of viscous terms are taken to be zero, 
i.e.; $\Phi(\tau_0)=0$ and $\Pi(\tau_0)=0$. 
We use the \textit{non-ideal} EoS ($\varepsilon-3P\neq 0$) 
obtained from equations (\ref{e3pt4} $\&$ \ref{pt4}) 
to close the system. We take critical temperature $T_c$ to be 190 MeV. 
We will not vary the height or width (controlled by the parameter 
$a$ and $\Delta T$ respectively) of the $\zeta/s$ curve in this analysis. These parameters are kept to their 
base values: $a=0.901$ and $\Delta T=T_c/14.5$ throughout this analysis.

By numerically solving the hydrodynamical equations describing the longitudinal expansion 
of the plasma (\ref{evo1}-\ref{evo3}), we get the temporal evolution profile for $T(\tau)$, $\Phi(\tau)$ 
and $\Pi(\tau)$. We can evolve the hydrodynamics till the temperature of the system reaches critical temperature, i.e.; $\tau_f$. 

In order to compare the effect of \textit{non-ideal} EoS on hydrodynamical evolution and dilepton yields we will compare these 
results with \textit{ideal} EoS ($\varepsilon-3P= 0$) of a gas of massless quarks and gluons. In this case, within Bjorken flow 
one can solve hydrodynamical evolution equations analytically to obtain the temperature profile as 
$T=T_0~(\frac{\tau_0}{\tau})^{\frac{1}{3}}$\cite{Bjorken:1982qr}. We have studied the temperature profile for both EoSs in our 
previous work and we found that hydrodynamical evolution gets significantly slowed down in case of 
\textit{non-ideal} EoS and system spends more time near $T_c$\cite{Bhatt:2010cy}.

Next we include the another \textit{non-ideal} effect, viscosity in the calculations.  
Now we study the longitudinal pressure $P_{z} = P + \Pi - \Phi$ of the system. 
It is already seen that in such a scenario, the viscous contribution to the equilibrium pressure makes the 
effective longitudinal pressure of the system zero, triggering \textit{cavitation}\cite{Rajagopal:2009yw}.
Hydrodynamics is applicable only till $\tau_c$ in case of occurrence of cavitation instead of $\tau_f$. 
This calculation is presented in our previous work\cite{Bhatt:2010cy} in detail and will not be repeated here. 
We quote the final results here: Eventhough system reaches $T_c$ at $\tau_f=\rm{5.5}~fm/c$ only, much before that 
at $\tau_c=\rm{2.5} ~fm/c$ it undergoes cavitation at a temperature 210 MeV.

Once we get the temperature profile we can calculate the desired dilepton yields as discussed in Section \ref{dilepton yields}. 
We again emphasise that we must be integrating the rates from $\tau_0$ to $\tau_1=\tau_c$ instead of $\tau_1=\tau_f$ in the 
case of cavitation, to avoid over-estimation of the yields\cite{Bhatt:2010cy}. From equations (\ref{tot-yield1}-\ref{tot-yield2}) 
we can now calculate the dilepton yields as functions of invariant mass $M$, transverse momentum perpendicular 
to collision axis $p_T$ and rapidity $y$ of the dileptons. We present all our calculations at the mid rapidity 
region of the dileptons ($y=0$).


Fig.[\ref{fig:2}] shows the dilepton yield $dN/dMdy$ calculated using \textit{ideal} (massless) and \textit{non-ideal} EoS. Effects 
of viscosity (both in hydrodynamics and distribution function) are ignored. 
In this calculation we take $0.5 GeV<p_{T}<2 GeV$. 
From the figure it is clear that \textit{non-ideal} EoS yields significantly larger dilepton flux
as compared to the \textit{ideal} EoS. At $M=1$ GeV, dilepton flux from the \textit{non-ideal} EoS is about 125$\%$ larger  
than that from the \textit{ideal} EoS case. This behavior can be understood by the fact that system cools slowly in the case of 
\textit{non-ideal} EoS allowing a higher temperature over a longer period, compared to \textit{ideal} EoS. It takes almost double 
the time for \textit{non-ideal} EoS to reach $T_c$. Since rates are dependent on temperature and an integration 
over $\tau_0$ and $\tau_1=\tau_f$, more dileptons are produced in the case of \textit{non-ideal} EoS.

It must be noted here that while calculating the particle spectra we use $\tau_1=\tau_f$ as we have cavitation in the system. 
As we demonstrated in our previous work\cite{Bhatt:2010cy} particle rates should be integrated upto 
$\tau_c$ and if we include $\tau_f$ instead of $\tau_c$ we will end up having a large over-estimation. 
In what follows we are presenting the correct particle yields by taking into consideration of the effect of 
cavitation. 
In Fig.[\ref{fig:3}] we show the dilepton yield at a fixed low invariant mass $M=0.525~GeV$ as a function of 
transverse momentum $p_T$ of the dileptons. Here we show the effect 
of inclusion of viscous corrections to the distribution function and dilepton rates separately. 
The solid curve shows the case without any 
viscous corrections to the rates (equation (\ref{dilrate0})): $\delta f=0$. 
Now inclusion of shear viscosity corrections to the distribution function (denoted by $\delta f=\delta f_{\eta}$) 
makes an increase in the dilepton production especially at the higher $p_T$. This result is in accordance with that of 
Ref.\cite{Dusling:2008xj}. When we consider only the bulk viscosity corrections ($\delta f=\delta f_{\zeta}$) we can see 
that the spectra gets suppressed at the high $p_T$ regime. 
This is result is in accordance with Refs.\cite{Monnai:2009ad,Bhatt:2010cy} 
where it is shown that effect of bulk viscosity is to suppress particles with high $p_T$. 
We can see that effect of shear and bulk viscosity oppose each other and the total contribution is represented 
in the curve ($\delta f_{\eta}+\delta f_{\zeta}$). 
As $p_T$ is increased the corrections are becoming high and the condition $f_0>>\delta f$ may be getting violated. 

\begin{figure}
\includegraphics[width=8.6cm]{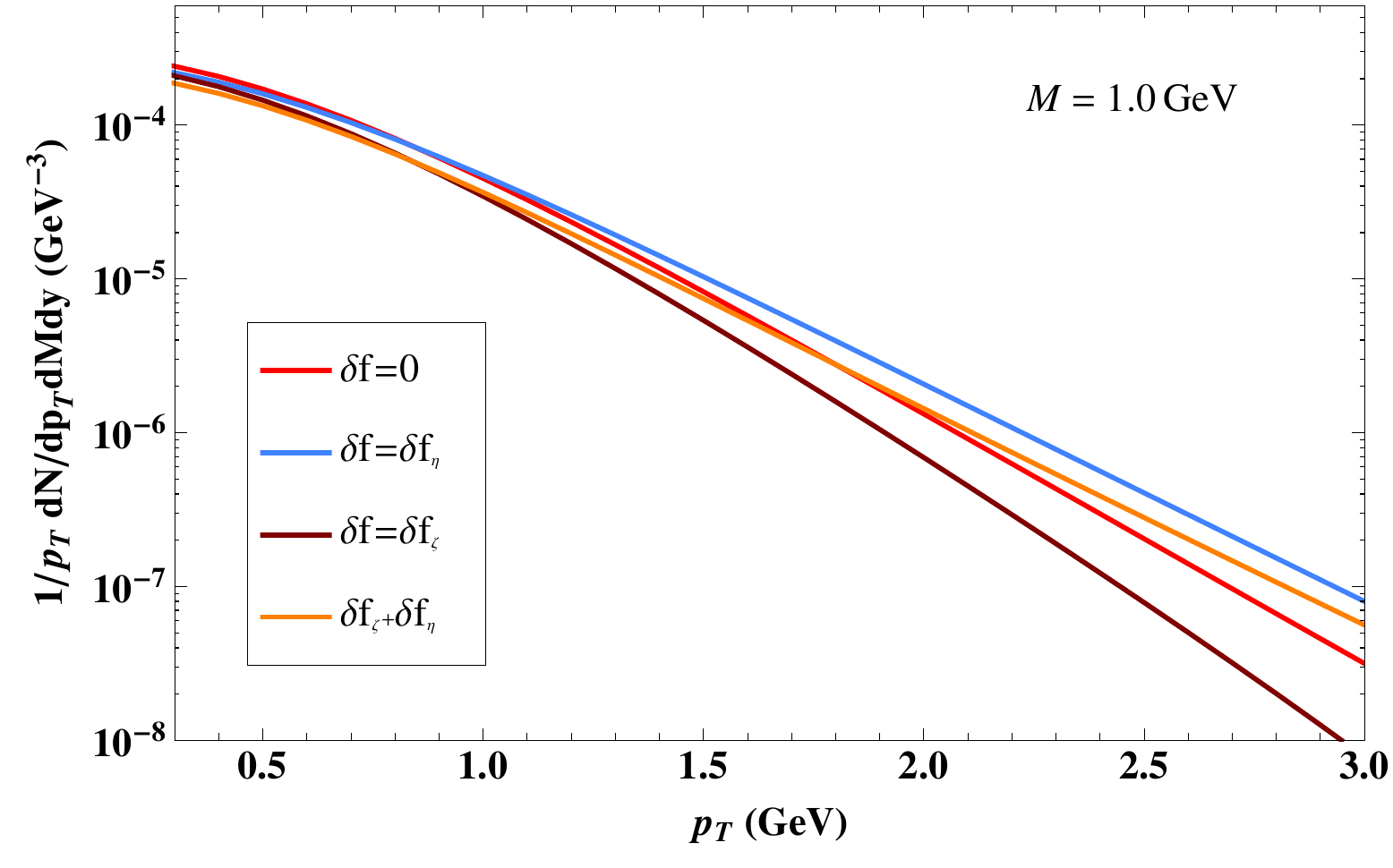}
\caption{Same as in Fig.[\ref{fig:3}], but for invariant mass $M=1.0~GeV$.}\label{fig:4}
\end{figure}
Next we consider the same dilepton rate as in previous case, calculated for a high invariant mass $M=1.0~GeV$. 
The results are shown in Fig.[\ref{fig:4}]. 
The solid curve shows the $\delta f=0$ case and  $\delta f=\delta f_{\eta}$ case is increasing the spectra as expected, however 
corrections to the spectra are becoming increasingly high with high $p_T$. 
One can see that  inclusion of modifications due to bulk viscosity alone heavily suppresses the dilepton spectra in this case 
($\delta f=\delta f_{\zeta}$). Validity of viscous corrections are under scrutiny here as values of both $p_T$ and $M$ are high 
making the corrections diverging. Unlike the case for photons, here corrections to the dilepton rates are dependent upon $M$ also 
(equations (\ref{dlratefinal}, \ref{visc-fcts})). 
Our results show that as invariant mass $M$ increases the viscous corrections become large even at smaller $p_T$ and thereby violating the 
condition $f_0>\delta f$. So the validity of the assumption $f_0>\delta f$ depends on $p_T$ as well as $M$ unlike the case of thermal photons 
where only $p_T$ dependence was there.

\begin{figure}
\includegraphics[width=8.6cm]{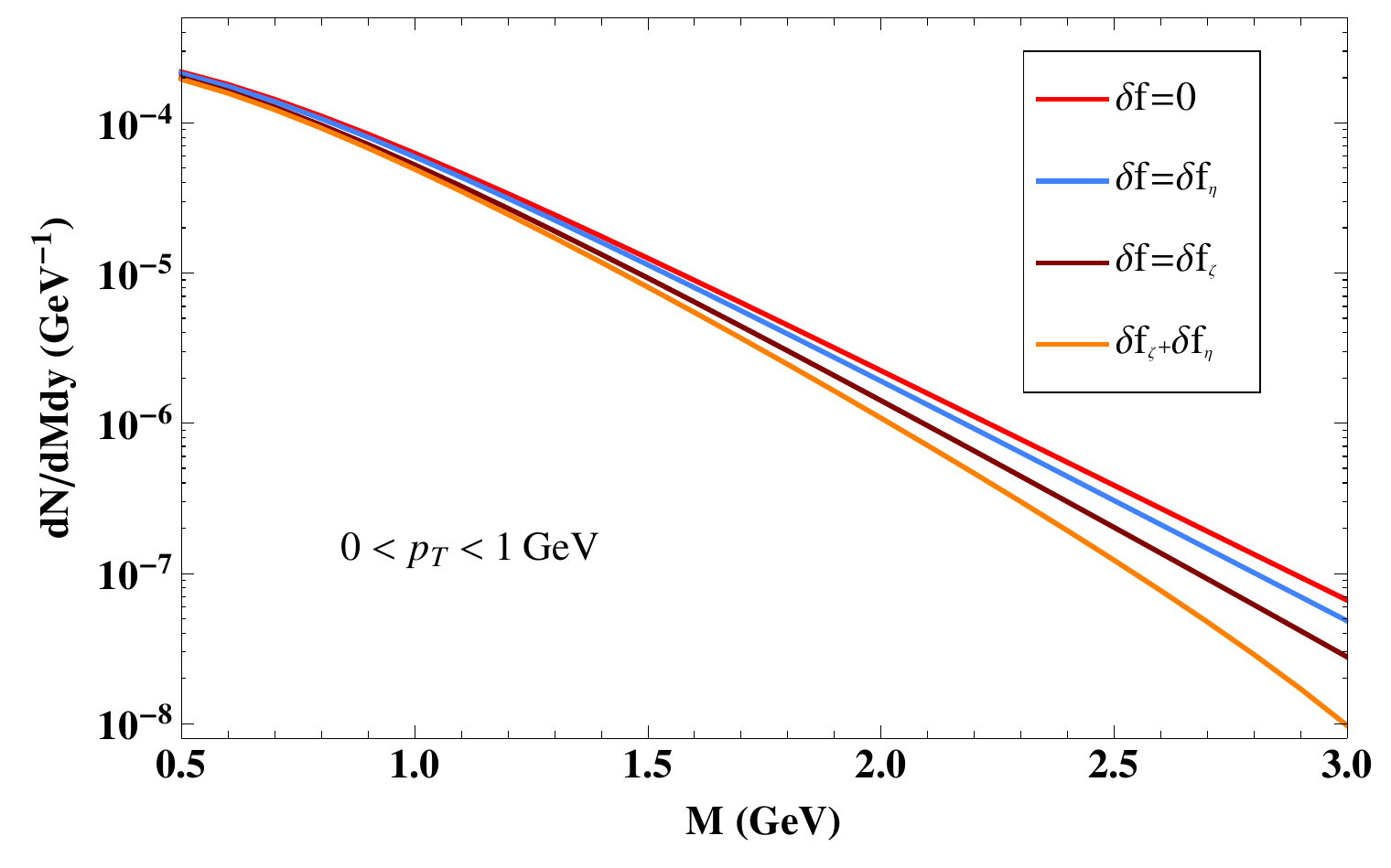}
\caption{Invariant mass distribution of mid-rapidity thermal dileptons in RHIC scenario calculated at the low $p_T$ regime. 
Effect of viscous corrections to the rate are shown separately.}\label{fig:5}
\end{figure}

Finally we plot the dilepton rate $dN/dMdy$ as a function of invariant mass of the dileptons in the low $p_T$ regime in 
Fig.[\ref{fig:5}]. We have used 
$p_{T_{min}}=\rm{0~} GeV$ and $p_{T_{max}}=\rm{1~} GeV$ in equation (\ref{tot-yield2}) while calculating the spectra. 
It is clear that approximation used for finding out viscous modifications in rates are more valid at the regime where both $M$ and $p_T$ 
values are small.

\section{SUMMARY}

We have calculated the first order corrections arising due to bulk and shear viscosities in 
dilepton production rates from $q\bar q$ annihilation. Viscous corrections to the distribution 
functions are obtained using the 
14-moment Grad's method. We have studied the role of \textit{non-ideal} effects near $T_c$ 
due to bulk viscosity, EoS and 
viscosity induced cavitation on the thermal dilepton production. 
Minimal value for  shear viscosity $\eta/s=1/4\pi$  is used in in this analysis.
We have shown that \textit{non-ideal} EoS taken from lattice results alone can significantly 
enhance the thermal 
dilepton production. Viscosities also enhance the particle production, however 
bulk viscosity can induce cavitation much before system reaches critical temperature making hydrodynamical 
description invalid. Thus role of bulk viscosity cannot be neglected in particle productions in heavy ion collisions. 
Finally the viscous corrections in distribution function grows in uncontrolled ways 
if the momentum and/or invariant mass of the dilepton increases and thus the 14-moment Grad's method breaks down.

\acknowledgements 
Authors would like to thank K. Dusling for providing his 3-dimensional hydrodynamical 
calculation results. One of the authors (SKV) would like to thank R. Bhalerao and R. Rapp for discussions.

\end{document}